# Agentic AI Enhances Physician Trust in Clinical Decision Making

**Zhiling Yan, MS[1], Zhe Fang, MD, PhD[2], David J King, MD[3], Ann Pongsakul, MD[4], Eashan Adhikarla, PhD[1], Hui Ren, MD, PhD[5], Sunyang Fu, PhD[6], Quanzheng Li, PhD[5], Lifang He, PhD[1], Xiang Li, PhD[5], Hongfang Liu, PhD[6], Yonghui Wu, PhD[7], Lichao Sun, PhD[1*]**

**[1]Department of Computer Science and Engineering, Lehigh University, Bethlehem, PA, USA; [2]Department of Epidemiology, Harvard T.H. Chan School of Public Health, Boston, MA, USA; [3]Department of Medicine, Division of Cardiology, University of Florida, Gainesville, FL, USA; [4]McGovern Medical School, University of Texas Health Science Center at Houston, Houston, TX, USA; [5]Department of Radiology, Massachusetts General Hospital and Harvard Medical School, Boston, MA, USA; [6]McWilliams School of Biomedical Informatics, University of Texas Health Science Center at Houston, Houston, TX, USA; [7]Department of Health Outcomes & Biomedical Informatics, College of Medicine, University of Florida, Gainesville, FL, USA**

**Abstract**

*Medical AI has shifted from reasoning to agentic AI, a new paradigm that autonomously invokes external tools during reasoning, rendering intermediate reasoning steps and tool outputs transparent to users. Although proven to outperform previous models, physician trust in agentic AI remains largely unexplored. To address this, three physicians evaluated 315 multimodal clinical cases quantifying both process-oriented cognitive trust and outcome-oriented behavioral reliance. Comparing agentic AI against non-agentic baselines, physicians exhibited significantly higher cognitive and behavioral trust for the agentic model ($P < 0.001$). Specifically, on treatment planning tasks, physicians trusted the agentic reasoning most, preferring it in 89.57% of cases. Furthermore, process-oriented cognitive trust is significantly associated with outcome-oriented behavioral reliance ($P < 0.001$). However, measurable over-reliance on incorrect agentic outputs still exists, highlighting the inherent limitations of decision-logic transparency alone and underscoring the continuous need for rigorous clinician oversight.*

**Introduction**

Artificial intelligence (AI) highlighted by large language models (LLMs) has been progressing rapidly in the medical domain.[1] Whereas early AI systems primarily functioned as enhanced search engines providing direct answers to simple clinical queries, the current paradigm has shifted toward reasoning AI.[2] These sophisticated models are capable of generating explicit intermediate reasoning steps before arriving at a conclusion, enabling them to tackle complex, real-world clinical challenges.[3] These breakthroughs suggest that AI has promising potential to enhance patient care, yet physician trust remains a critical barrier to its adoption in real-world healthcare.[4]

Physicians are more likely to trust an AI system if its underlying reasoning logic is explainable and clinically sound.[5] While reasoning AI models have evolved to generate step-by-step rationales, these rationales are self-reported and not anchored to verifiable evidence. It means they fail to provide traceable evidence to corroborate their logic. When physicians cannot verify the source and validity of the AI's logic, their trust is compromised, diminishing their willingness to employ AI in real-world clinical setting. Recent advances in “agentic AI” offer a qualitatively different and promising paradigm to bridge this crucial trust gap. Unlike previous reasoning AI merely narrating its reasoning, agentic AI is characterized by its capacity to autonomously invoke specialized external tools (e.g., clinical calculators, image analysis modules, and up-to-date literature search) during its reasoning phase.[6] Crucially, the intermediate execution steps and the outputs of these tools remain visible to the user. This shifts the basis of trust from believing an opaque model to inspecting traceable, externally grounded evidence.

Despite this promise, the impact of agentic AI on physician trust remains largely unexamined. Existing studies on agentic systems have predominantly focused on benchmarking clinical accuracy,[7] implicitly treating a correct answer as sufficient for adoption. Yet trust is a multidimensional construct not captured by accuracy alone. It spans both cognitive trust in the reasoning process and behavioral trust expressed through reliance on the final decision.[8, 9] Evaluating whether agentic AI earns appropriate physician trust therefore requires a framework that measures these dimensions directly, rather than inferring them from performance metrics.

To address this critical gap, this study systematically investigates whether and to what extent agentic AI enhances physician trust in clinical decision-making compared to previous non-agentic AI. Moving beyond simplistic performance metrics, we establish a clinical trust ranking framework to quantify physician trust across two distinct dimensions, i.e., process-oriented cognitive trust and outcome-oriented behavioral trust. We curated 315 multimodal, free-response clinical cases, spanning diagnoses, treatments, and lab tests, and engaged three physicians from three independent healthcare systems to systematically rank intermediate reasoning traces and final generated decisions of the models. We further analyze case-level factors associated with physician confidence versus distrust, and examine the gap between physicians' subjective reliance and the models' objective accuracy to assess potential over-reliance.

## Method

In this section, we introduce the overall pipeline of this study, including clinical data curation, experimental setup, and the evaluation framework (**Figure 1**). The code is available at https://github.com/ZhilingYan/AgenticBench.

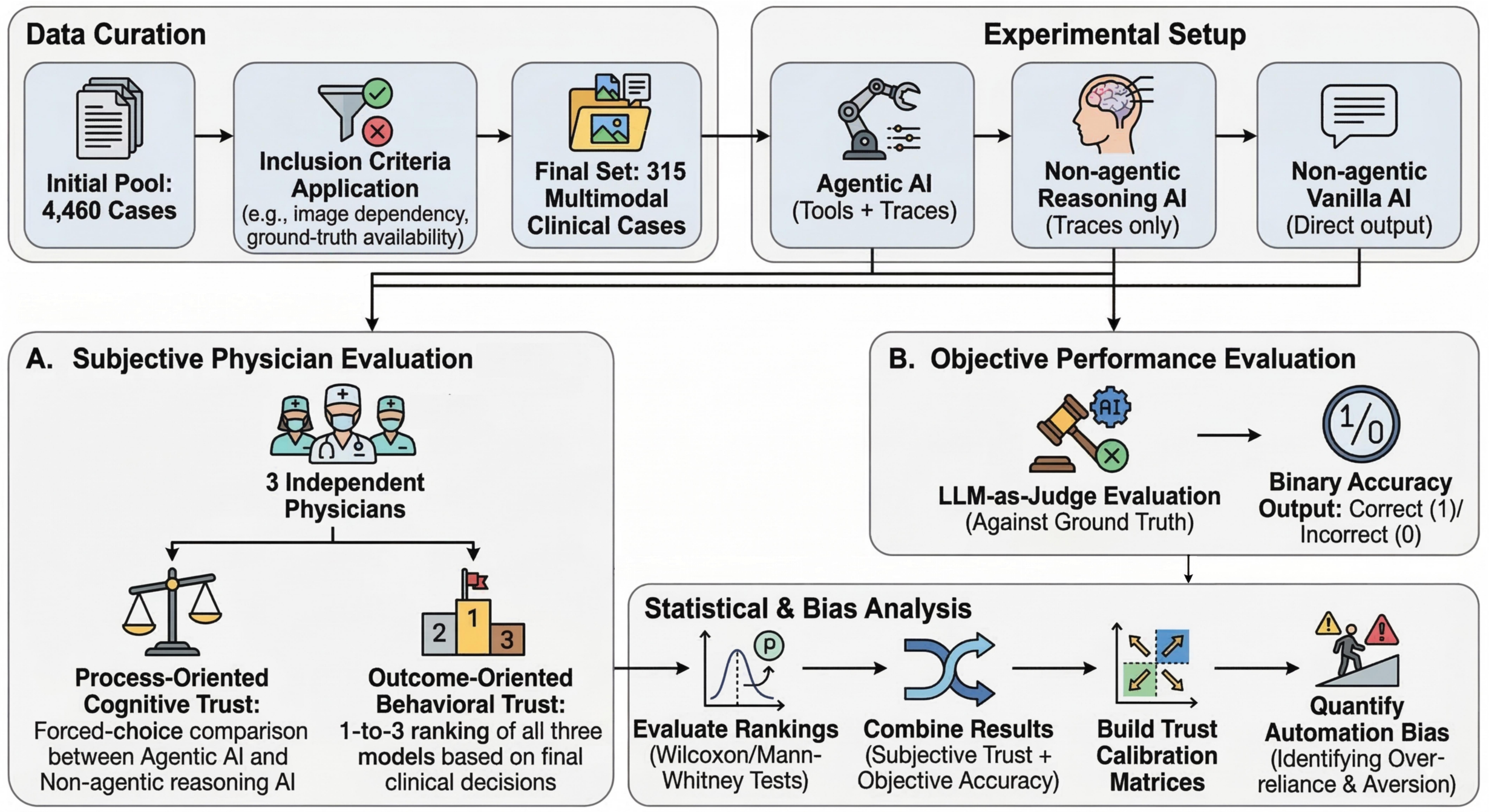

**Figure 1.** Overall study design and evaluation pipeline. The workflow illustrates the data curation process, the experimental setup of three AI models, the dual-track evaluation (subjective physician trust vs. objective AI accuracy), and the subsequent statistical analysis to quantify physician trust and behavioral automation bias.

### Data curation

To ground the study in a representative set of realistic cases, we started from an initial pool of 4,460 cases.[10] Sources included standardized examinations and textbooks including United States Medical Licensing Examination (USMLE), Comprehensive Osteopathic Medical Licensing Examination (COMLEX-USA), 17 American specialty board exams and image-rich sources such as the NEJM Image Challenge.

To evaluate performance in multimodal clinical settings, all cases were manually screened by an annotator and retained for further analysis if they met the following inclusion criteria: (1) images are required to inform clinical decision; (2) the case is directly relevant to the study’s clinical reasoning tasks; (3) a single and definitive ground-truth answer exists; and (4) the AI provides eligible decision-support answers without safety concerns or refusals to answer. This process resulted in a final set of 315 cases that met these criteria. **Table 1** summarizes the distribution of included cases by clinical decision-making task and body system.

**Table 1.** Overview of clinical cases. Number of cases per body system and clinical decision-making tasks.

| | All | Lab test | Diagnosis | Treatment |
|---|---|---|---|---|
| n | 315 | 42 (13.3%) | 158 (50.2%) | 115 (36.5%) |
| ***Body system*** | | | | |
| Cardiovascular | 67 (21.3%) | 8 (2.5%) | 31 (9.8%) | 28 (8.9%) |
| Skeletal | 62 (19.7%) | 12 (3.8%) | 23 (7.3%) | 27 (8.6%) |
| Respiratory | 46 (14.6%) | 3 (1.0%) | 27 (8.6%) | 16 (5.1%) |
| Nervous | 34 (10.8%) | 3 (1.0%) | 19 (6.0%) | 12 (3.8%) |
| Digestive | 34 (10.8%) | 4 (1.3%) | 25 (7.9%) | 5 (1.6%) |
| Integumentary | 20 (6.4%) | 5 (1.6%) | 6 (1.9%) | 9 (2.9%) |
| Reproductive | 19 (6.0%) | 3 (1.0%) | 9 (2.9%) | 7 (2.2%) |
| Endocrine | 13 (4.1%) | 2 (0.6%) | 9 (2.9%) | 2 (0.6%) |
| Lymphatic | 8 (2.5%) | 0 (0%) | 3 (1.0%) | 5 (1.6%) |
| Muscular | 8 (2.5%) | 2 (0.6%) | 4 (1.3%) | 2 (0.6%) |
| Urinary | 4 (1.3%) | 0 (0%) | 2 (0.6%) | 2 (0.6%) |

**Workflow for Quantifying Physician Trust**

To systematically evaluate physician trust, assessments were performed under blinded review by three physicians from three independent healthcare institutions. The cases were randomly assigned and reviewed independently under blinding. To quantify physician trust, we moved beyond simplistic accuracy metrics and established a clinical trust ranking framework, encompassing two distinct dimensions, i.e., process-oriented cognitive trust and outcome-oriented behavioral trust.

***Process-Oriented Cognitive Trust***. It reflects the physician's subjective evaluation of the AI's competence and reliability based on its reasoning logic. Physicians were presented with the step-by-step reasoning traces of two models (the agentic AI versus the non-agentic reasoning AI) under double-blind conditions. They were strictly instructed to evaluate and select which reasoning process was superior based on the following dimensions:

- Logical Flow: Which model's clinical reasoning logic is more coherent and closely aligns with my diagnostic deduction process?
- Evidence Grounding: Which model more accurately anchors its reasoning on key clinical clues (e.g., abnormal metrics, medical history) without generating hallucinations?
- Transparency & Explainability: Which model's reasoning trace provides a clearer understanding of how it arrived at the final clinical decision?

After evaluating these dimensions, physicians provided an overall comparative judgment to answer the core prompt: Based on presented reasoning traces, which model's professional competence am I willing to trust? The vanilla AI baseline was deliberately excluded from this specific track. Because it generates direct answers without reasoning.

***Outcome-Oriented Behavioral Trust***. Behavioral trust is the objective manifestation of reliance, i.e., the physician's actual behavioral intent to adopt the AI's output in a high-stakes environment. Physicians evaluated the final clinical decisions (e.g., diagnoses, treatment plans) generated by all three models (agentic AI and two baselines, i.e., reasoning AI and vanilla AI). They ranked the outputs strictly from 1 (Highest Reliance Intention / Best) to 3 (Rejection / Worst) based on the following dimensions:
- Clinical Reliance: In a real-world clinical setting, I would be willing to directly adopt this model's recommendation without needing to extensively cross-reference literature.
- Risk Assumption: If I were to execute the clinical plan recommended by this model, I feel confident in its safety for the patient.
- Switching Behavior: If my initial clinical judgment differed from this model's recommendation, this model's output is persuasive and comprehensive enough to prompt me to revise my original decision.
Physicians then provided an overall ranking, ordering the outputs strictly from 1 (Highest Reliance Intention / Best) to 3 (Rejection / Worst) to answer the core prompt: Based on the presented output, which model's clinical decisions am I willing to adopt?

**Model Selection**
To investigate the impact of the agentic AI paradigm on physician trust, we selected a representative set of state-of-the-art large language models that represent different stages of AI evolution.
- Agentic AI: GPT-o3[11] was chosen as the primary representative of agentic AI. Unlike traditional models, GPT-o3 autonomously invokes specialized external tools, such as image analysis, clinical calculators, and literature search, and exposes these intermediate reasoning and invocation traces to the user.
- Non-agentic reasoning AI: GPT-o3*[11] was chosen as the primary representative of reasoning AI. This is the exact same GPT-o3 model but with its agentic tool-use capabilities disabled. It serves as a strict ablation baseline, representing a highly capable reasoning AI that generates step-by-step rationales but cannot gather verifiable external evidence.
- Non-agentic vanilla AI: GPT-4o[12] was chosen as the primary representative of vanilla AI. This model represents the previous paradigm of standard AI, which generates direct clinical outputs without reasoning.
While we acknowledge that newer models continue to emerge, GPT-o3 serves as an ideal exemplar for this study because it explicitly displays the specific agents it invokes, allowing for a granular analysis of how tool transparency affects physician trust.

**Querying the LLMs**
To ensure consistency, the inputs, including the patient narrative (e.g., medical history, current symptoms), related multimodal images, and the specific clinical queries, were kept strictly identical across all three models. For all queries, we appended the instruction "Name up to five answers" to standardize the output format.[13]
- Agentic AI (GPT-o3): We queried GPT-o3 through its official web interface (https://chatgpt.com/). This approach was necessary because the complete, step-by-step reasoning traces, including the granular details of agent invocations, were only accessible to users via the website interface.
- Baselines (GPT-o3* and GPT-4o): We utilized the official API for the two non-agentic baselines. For GPT-o3*, the specific parameter that enables agent invocation was configured to be disabled in the API request to ensure it functioned purely as a reasoning model.
To ensure a rigorous comparison, we emphasize that the underlying model architecture and base computational capabilities remain identical across both the web interface and the API. The choice of platform solely dictates the user-facing visibility of the intermediate tool-use traces and introduces no performance or evaluation bias. All queries and data collection were executed between June 1st and June 16th, 2025.

**Statistical evaluation**
To mitigate the central tendency bias inherent in traditional Likert scales, our evaluation utilized 2-point and 3-point ranking paradigms.[14] Because this approach generates strict ordinal data reflecting relative physician preference, we applied non-parametric statistical tests. Specifically, to evaluate significant pairwise differences in trust rankings between models (e.g.., GPT-o3 versus GPT-o3*, and GPT-o3 versus GPT-4o), a one-sided Wilcoxon signed-rank test was employed, with Bonferroni correction applied to control for multiple testing. For unpaired data comparisons, we utilized the one-sided Mann-Whitney U test.

Another critical objective of our statistical evaluation was to investigate the gap between subjective physician trust and objective AI accuracy, a phenomenon known as automation bias.[15] To measure automation bias, we constructed trust calibration matrices. Cases where a model's output was objectively incorrect but still ranked as highly trusted (Rank 1) by physicians were categorically defined as instances of over-reliance. Conversely, cases where the output was correct but ranked low were defined as algorithmic aversion. We statistically compared the prevalence of over-reliance and algorithmic aversion between agentic and non-agentic models using Chi-square tests. All statistical analyses were conducted using Python 3.12.6, leveraging the scipy package for hypothesis testing ($\alpha = 0.05$).

**Performance Evalution**

To evaluate the objective clinical accuracy of the models' final outputs and establish a baseline for our over-reliance analysis, we employed an LLM-as-judge evaluation paradigm.[16] Specifically, we utilized Gemini 3 Flash[17] to evaluate whether each model's generated clinical decision (e.g., diagnosis, treatment plan, or lab-test recommendation) strictly contained the clinical ground-truth answer of each case. To ensure strict objectivity, the judge model was constrained to only deterministic semantic verification rather than open-ended grading. It was programmed solely for binary semantic verification: If the model's output included the ground-truth answer, it was scored as correct (1); if not, it was scored as incorrect (0). By establishing this objective accuracy, we were subsequently able to cross-reference these performance outcomes with the physicians' subjective behavioral trust rankings.

**Evaluation and Results**

***Agentic AI achieves higher process-oriented cognitive trust.*** To evaluate process-oriented cognitive trust, physicians compared the intermediate reasoning traces of the agentic AI (GPT-o3) against the reasoning AI (GPT-o3*). For each case, physicians judged whether the agentic AI's reasoning trace earned their trust. "Preferred" indicates the agentic trace was judged trustworthy/more reliable, "Rejected" indicates it was not. In fact, physicians could reject both reasoning traces, so a case not counted as preferring the agentic AI does not necessarily favor the reasoning AI (GPT-o3*). Results in **Table 2** show that physicians exhibited a higher cognitive trust preference for the agentic AI, ranking its reasoning traces as more reliable in 84.13% of the 315 multimodal cases (95% CI 0.8009–0.8816). Notably, the remaining 15.87% of cases reflect a lack of preference for the agentic AI's reasoning trace and do not imply that the reasoning AI's trace was preferred instead. For the three clinical reasoning tasks related to lab tests, diagnoses, and treatments, The trust preference was 88.10% for lab-test recommendations (n = 42; 95% CI 0.7830–0.9789), 79.11% for diagnosis (n = 158; 95% CI 0.7278–0.8545), and 89.57% for treatment planning (n = 115; 95% CI 0.8398–0.9515). Overall, the transparent display of tool use and verifiable evidence garnered significantly higher physician trust compared to black-box reasoning ($P < 0.001$).

**Table 2.** Physician-evaluated process-oriented cognitive trust of agentic AI reasoning traces across clinical tasks.

| Clinical Task | Agentic AI Rejected (Low Cognitive Trust) | Agentic AI Preferred (High Cognitive Trust) |
|---|---|---|
| Overall (n = 315) | 50 (15.9%) | 265 (84.1%) |
| Lab test (n = 42) | 5 (11.9%) | 37 (88.1%) |
| Diagnosis (n = 158) | 33 (20.9%) | 125 (79.1%) |
| Treatment (n = 115) | 12 (10.4%) | 103 (89.6%) |

As illustrated in **Figure 2**, the agentic model searched external guidelines, caught the safety pitfall its initial answer had missed, and corrected its plan to the right recommendation. Both its reasoning steps and the supporting evidence, i.e., the retrieved guideline, stayed visible to the physician, making the decision verifiable rather than opaque.

***Agentic AI attains superior outcome-oriented behavioral trust.*** Beyond the reasoning process, we assessed outcome-oriented behavioral trust by analyzing physicians' actual behavioral intent to adopt the final clinical decisions (e.g., executing the generated treatment plan or diagnosis) using the 3-point ranking scale (where higher scores indicate greater reliance intention). Agentic AI achieved the highest average physician trust rankings compared to all baselines across all tasks, as shown in **Table 3**. Specifically, the agentic AI (GPT-o3) outperformes both the non-agentic reasoning AI (GPT-o3*; $P < 0.001$, rank-biserial correlation $r_{rb} = 0.20$) and the vanilla non-agentic baseline (GPT-4o; $P < 0.001$, $r_{rb} = 0.38$).

**Case.** A 24-year-old woman arrives at the emergency department with an hour-long episode of palpitations. ... Her medical history includes poorly controlled asthma despite regular inhaler use. ... On examination, ... include a blood pressure of 104/70 mm Hg, a pulse of 194/min (regular), and a respiratory rate of 18/min. An ECG is provided [*Image omitted for brevity*]. Carotid sinus massage for 5 to 10 seconds does not resolve her palpitations.
*Question:* what is the next best step? (*Reference answer: verapamil.*)

**Non-agentic AI.** "The patient has a one-hour history of palpitations, and the ECG shows a narrow-complex regular tachycardia, likely AVNRT or AVRT. Based on standard protocols, adenosine is the first-line treatment. Final decision: **Administer IV adenosine 6–12 mg.**"

**Agentic AI.** (1) **Task Planning:** Analyzes the clinical vignette and sets the goal; (2) **Initial Assessment:** Interprets the ECG: "The rhythm is a narrow-complex regular tachycardia at 194 bpm, consistent with SVT. Standard first-line therapy is adenosine."; (3) **Self-Reflection & Safety Check:** Cross-references the proposed treatment with the patient's full medical history. Raises a critical safety flag: "The patient has a history of poorly controlled asthma. Is adenosine safe in this specific context?"; (4) **External Knowledge Retrieval:** Executes a targeted search of medical guidelines (e.g., query: "adenosine safety in asthma SVT"). Retrieves evidence indicating that adenosine is strictly contraindicated in patients with hyperreactive airway disease due to the severe risk of bronchospasm.; (5) **Correction & Final Decision:** Citing the evidence (*2015 ACC/AHA/HRS Guideline for the Management of Adult Patients With Supraventricular Tachycardia*), the model overrides its initial assumption and revises the plan: "Given the strict contraindication of adenosine in asthmatic patients, verapamil is better. Final decision: **Administer Verapamil 5–10 mg IV over ≥2 min and monitor blood pressure.**"

*Note: AI outputs have been condensed to highlight the core reasoning pathways.

**Figure 2.** Sample reasoning narrative: agentic vs. non-agentic model on the same clinical case.

**Table 3.** Physician-evaluated outcome-oriented behavioral trust rankings and statistical comparisons against the agentic AI (GPT-o3) across clinical tasks.

| **Model** | | **Rank 3 (Lowest Reliance)** | **Rank 2 (Middle Reliance)** | **Rank 1 (Highest Reliance)** | **Comparison vs. GPT-o3** | |
|---|---|---|---|---|---|---|
| | | | | | P-value (adjusted) | rank-biserial correlation $r_{rb}$ |
| All (n = 315) | GPT-o3 | 60 (19.0%) | 101 (32.1%) | **154 (48.9%)** | - | - |
| | GPT-o3* | 97 (30.1%) | **117 (37.1%)** | 101 (32.1%) | **<0.001** | 0.20 |
| | GPT-4o | **158 (50.2%)** | 97 (30.1%) | 60 (19.0%) | **<0.001** | 0.38 |
| Task Category | | | | | | |
| Lab test (n = 42) | GPT-o3 | 8 (19.0%) | 16 (38.1%) | **18 (42.9%)** | - | - |
| | GPT-o3* | 14 (33.3%) | **15 (35.7%)** | 13 (31.0%) | 0.2461 | 0.10 |
| | GPT-4o | **20 (47.6%)** | 11 (26.2%) | 11 (26.2%) | 0.0932 | 0.38 |
| Diagnosis (n = 158) | GPT-o3 | 35 (22.2%) | 49 (31.0%) | **74 (46.8%)** | - | - |
| | GPT-o3* | 43 (27.2%) | **61 (38.6%)** | 54 (34.2%) | 0.1779 | 0.20 |
| | GPT-4o | **80 (50.6%)** | 48 (30.4%) | 30 (19.0%) | **<0.001** | 0.32 |
| Treatment (n = 115) | GPT-o3 | 19 (16.5%) | 36 (31.3%) | **60 (52.2%)** | - | - |
| | GPT-o3* | **40 (34.8%)** | **40 (34.8%)** | 35 (30.4%) | **0.0029** | 0.23 |
| | GPT-4o | **56 (48.7%)** | 39 (33.9%) | 20 (17.4%) | **<0.001** | 0.48 |

Pairwise comparisons further emphasize this behavioral reliance shift at the task level. For treatment planning tasks, physicians were significantly more willing to adopt the actionable plans generated by GPT-o3 over those from GPT-o3* ($P = 0.0029$, $r_{rb} = 0.23$) and GPT-4o ($P < 0.001$, $r_{rb} = 0.48$). A similar significant preference was observed in diagnosis tasks against GPT-4o ($P < 0.001$, $r_{rb} = 0.32$). These robust statistical findings validate that the integration of agentic capabilities not only improves subjective explanation satisfaction but actively drives physicians' behavioral intent to rely on the AI's final clinical output.

**Table 4.** Distribution of primary agentic behaviors driving physician cognitive trust (n = 265).

| Category | Explicit Literature & Guideline Citation | Visual Region-of-Interest (ROI) Grounding | Self-Correction | Clinical Calculator Invocation |
|---|---|---|---|---|
| # Cases (n = 265) | 119 (44.9%) | 69 (26.0%) | 66 (24.9%) | 11 (4.2%) |

***What skills make physicians trust agentic AI more?*** To understand the underlying drivers of the 84.13% (n = 265) cognitive trust preference, we conducted a qualitative thematic analysis of the physicians' evaluation notes alongside the models' tool invocation logs. We identified four primary agentic behaviors that significantly bolstered clinical confidence. **Table 4** summarizes the distribution of these trust-enhancing skills.

The most prominent driver of trust was explicit literature and guideline citation (44.9%, n = 119), as the model's ability to search and cite up-to-date medical guidelines. In multimodal contexts, visual Region-of-Interest (ROI) grounding (26.0%, n = 69) further reassured evaluators. It proved that the model was genuinely interpreting the visual data rather than hallucinating a diagnosis based solely on the text narrative. Furthermore, proactive self-correction (24.9%, n = 66) played a critical role in complex cases. When the agentic AI verified its initial hypotheses, recognized clinical contradictions (e.g., a drug contraindication due to patient history), and corrected its reasoning path, physicians highly valued this transparent self-reflection as a robust mechanism for ensuring patient safety. Finally, clinical calculator invocation (4.2%, n = 11) showed that the autonomous computation of quantitative clinical scores mirrored human workflows, reducing the physicians' cognitive load.

**Table 5.** Distribution of failure modes causing physician distrust in agentic AI reasoning (n = 50).

| Category | Circular Reasoning and Overthinking | Misapplication of Retrieved Evidence | Vision-Text Misalignment | Unresolved Tool Contradictions |
|---|---|---|---|---|
| # Cases (n = 50) | 21 (42.0%) | 15 (30.0%) | 10 (20.0%) | 4 (8.0%) |

***What reasons make physicians do not trust agentic AI?*** Conversely, in the 15.87% (n = 50) of cases where physicians rejected the agentic reasoning, we categorized these failure modes into four distinct patterns causing physician distrust, as shown in **Table 5**.

The most frequent cause of physician distrust was circular reasoning and overthinking (42.0%, n = 21). In these instances, the agent repeatedly invoked search tools or self-critique loops without advancing the diagnostic narrative, producing verbose, disjointed traces that physicians found unreliable and time-consuming to read. Another major detractor was the misapplication of retrieved evidence (30.0%, n = 15). Although the agent successfully accessed relevant literature, it occasionally misinterpreted nuanced clinical contexts. Furthermore, vision-text misalignment (20.0%, n = 10) undermined confidence in multimodal cases. When the image analysis agent missed critical visual cues or hallucinated findings, subsequent text-reasoning agents were inevitably led down incorrect paths. Finally, trust was occasionally broken by unresolved tool contradictions (8.0%, n = 4), where AI failed to synthesize or resolve conflicting information returned by different tools, such as a clinical calculator suggesting high risk alongside a guideline search advising conservative management.

***Process-oriented cognitive trust is significantly associated with outcome-oriented behavioral trust.*** To investigate whether trusting an AI's reasoning naturally leads to adopting its clinical plan, we analyzed the alignment between the physicians' cognitive preference and their behavioral ranking. We constructed a trust matrix (**Table 6**) to visualize this decision pathway across all 315 clinical cases. Statistical analysis demonstrated a strong, positive correlation (Spearman ρ = 0.608, $P < 0.001$ ), confirming that cognitive approval is highly related to behavioral reliance.

**Table 6.** Trust matrix shows the alignment between process-oriented cognitive trust and outcome-oriented behavioral reliance.

| Ranking | Rank 1 | Rank 2 | Rank 3 | Total |
|---|---|---|---|---|
| **Preferred** | 153 (57.7%) | 97 (36.6%) | 15 (5.7%) | 265 |
| **Rejected** | 1 (2.0%) | 4 (8.0%) | 45 (90.0%) | 50 |
| **Total** | 154 (48.9%) | 101 (32.1%) | 60 (19.0%) | 315 |

Specifically, among the 265 cases where physicians preferred GPT-o3's transparent reasoning trace (High Cognitive Trust), they assigned its final output the highest reliance intention (Rank 1) in 57.7% (n = 153) of the instances. This indicates that once trust is established through verifiable tool use, physicians are highly willing to act on the generated plan. In the 50 cases where physicians rejected the agentic reasoning due to circular logic or hallucinations, the subsequent behavioral trust plummeted, with 98.0% (n = 49) of the corresponding final outputs receiving a Rank 2 or Rank 3 (Rejection).

***Behavioral automation bias.*** To quantify behavioral automation bias, we evaluated the alignment between subjective physician trust and objective model accuracy. To construct the trust calibration matrices, we dichotomized the ordinal behavioral reliance rankings into a binary variable. Cases assigned Rank 1 (highest reliance) were defined as "Trust," while cases assigned Rank 2 or Rank 3 were aggregated as "Not-trust." Objective AI accuracy was strictly defined as "Correct" or "Incorrect" based on the ground truth answer. The resulting alignment distributions for both the agentic and non-agentic models are shown in **Figure 3**.

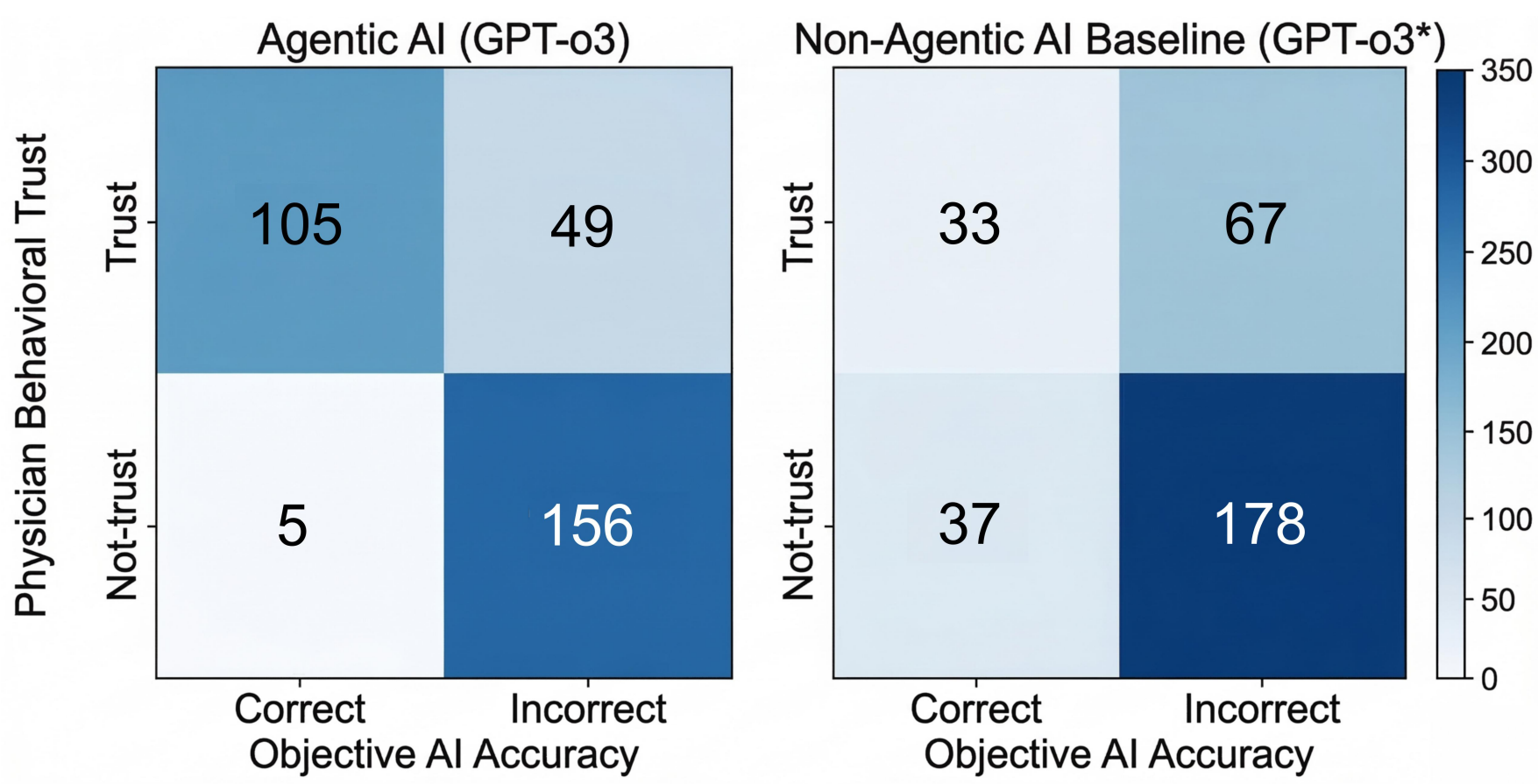

**Figure 3.** The trust calibration matrices visualize the intersection of objective AI accuracy and subjective physician behavioral trust for both the agentic AI (GPT-o3) and the non-agentic baseline (GPT-o3*).

Experiments demonstrate that physicians consistently exhibit over-reliance (trusting an incorrect output) and AI aversion (distrusting a correct output). For the agentic AI, the proportion of over-reliance is 23.9% (n = 49), and the proportion of AI aversion is 4.6% (n = 5). This phenomenon is notably more severe in the non-agentic reasoning AI, with the over-reliance proportion rising to 27.4% (n = 67) and the AI aversion proportion drastically increasing to 52.9% (n = 37). Through Chi-square testing on the overall trust calibration errors (combining both over-reliance and AI aversion cases), we found that although both problems exist across models, the agentic AI is significantly superior to the reasoning AI in mitigating these biases (Chi-square Testing $\chi^2 = 21.12$, $P < 0.001$).

**Discussion and Conclusion**

In this study, we established an evaluation framework to move beyond simplistic performance benchmarking, focusing instead on how the transition from reasoning AI to agentic AI impacts physician trust. Our findings demonstrate that the agentic paradigm enhances both process-oriented cognitive trust and outcome-oriented behavioral trust compared to non-agentic baselines ( $P < 0.001$ ). Qualitative feedback indicates that physicians attributed their increased trust to the transparent display of dynamic tool invocations. Behaviors such as explicit literature searches were perceived as providing a verifiable foundation that mirrors evidence-based clinical workflows.

Despite the high overall trust elicited by agentic AI, our findings reveal critical challenges that preclude its autonomous deployment. As evidenced by our trust calibration matrices, physicians still exhibited over-reliance in 23.9% of incorrect agentic cases. Furthermore, the models currently insufficient objective accuracy (34.9%) for clinical decision-making tasks indicates that agentic AI is still not mature enough to independently face patients.

A key contribution of our work is the use of multimodal, free-response cases with both clinical text and medical images. Earlier work has focused on text-only inputs or multiple-choice tasks, where current AI systems already perform well.[18] Multimodal, free-response reasoning tasks better stress the strengths of agentic reasoning, including cross-modal synthesis, evidence grounding, and tool use, which may explain why previous studies observed limited advantages for reasoning in simpler settings.[13]

We used a straightforward ranking scale across all tasks. While this enabled standardized scoring across physicians and specialties, it may be subject to personal bias. Furthermore, the expert evaluation cohort consisted exclusively of physicians practicing within the United States healthcare system. Given that clinical workflows, regulatory environments, and general attitudes toward AI can vary significantly across different medical cultures, the generalizability of our findings may be limited. In addition, objective accuracy was scored by an LLM judge. Although it is constrained to semantic alignment against a predefined ground-truth answer, this may still introduce potential bias. Future studies should explore broader clinician panels and more diverse clinical dimensions, such as impact on prognosis, guideline concordance, and equity considerations. Future studies should also assess broader socio-technical dimensions of trust, such as institutional accountability, liability, and workflow integration, beyond the individual physician–model level examined here.

While agentic AI exposes the execution and results of its tools (e.g., showing the retrieved search results), the underlying algorithmic decision-making process governing why, when, and how to invoke a specific tool remains a black box. Currently, this autonomous orchestration is largely uncontrollable by the clinician, which limits the user's ability to intervene or steer the AI’s reasoning trajectory proactively. Future studies should examine controllable and reliable use of agents, e.g., user-initiated authorization of agent calls, or sourcing from quality-checked medical databases. Strategies that could enable physicians to customize the types of agents to include in agentic AI models will be helpful. Future studies should also explore solutions to align agentic AI with regulations for clinical decision support, given the limited transparency of data inputs and training processes.

Our study utilized GPT-o3 rather than more recent iterations like GPT-5[19]. This choice was deliberately constrained by the models’ user interface designs, as newer models currently obfuscate specific agent invocations and intermediate traces from the user, rendering our granular analysis of tool-use transparency technically impossible. We recognize the inherent limitation of evaluating a comparatively older model in a rapidly evolving field. However, the core objective of this work is not to benchmark state-of-the-art accuracy, but rather to investigate whether agentic AI structurally enhances physician trust compared to non-agentic AI. In this context, GPT-o3 serves as a necessary and representative exemplar to study the psychological impact of visible tool use. Future work could also extend this framework to open-source multimodal reasoning models, which are typically smaller, more transparent, and less energy-intensive, and may achieve comparable performance without relying on closed commercial systems.

In conclusion, this study demonstrates that the paradigm shift from black-box reasoning to agentic AI fundamentally enhances physician trust in clinical decision-making. Through our evaluation of 315 multimodal cases, we established that the transparent, tool invocations of agentic AI significantly enhance both process-oriented cognitive trust and outcome-oriented behavioral reliance compared to non-agentic baselines. While this verifiable reasoning logic aligns closely with human clinical workflows, the persistent risk of automation bias highlights that transparency alone cannot entirely eliminate cognitive vulnerabilities. Ultimately, our findings advocate for integrating agentic reasoning into future healthcare AI across diagnostic, treatment, and lab-test domains. As models evolve to incorporate quality-checked medical sources, controllable agent execution, and robust clinician oversight, prospective clinical trials will be essential to determine how these enhanced trust dynamics translate into tangible improvements in patient care and clinical outcomes.

**Acknowledgment**
This study was partially supported by the National Science Foundation Grants CRII-2246067, ATD-2427915, NSF POSE-2346158, and NSF POSE-2449280, and grants from the Patient-Centered Outcomes Research Institute® (PCORI®) Award ME-2023C3-35934, the PARADIGM program awarded by the Advanced Research Projects Agency for Health (ARPA-H).